\documentclass[twocolumn,showpacs,preprintnumbers,amsmath,amssymb]{revtex4}

\usepackage{graphicx}
\usepackage{dcolumn}
\usepackage{bm}

\begin{document}


\title{ Quantum discord and noncontextual hidden variables models}

\author{Kazuo Fujikawa}
\affiliation{%
Institute of Quantum Science, College of Science and Technology,
Nihon University, Chiyoda-ku, Tokyo 101-8308, Japan
}%


\begin{abstract}
It is shown that theoretically viable  noncontextual hidden variables
models in $d=2$ lead to conflicting dispersion free 
expressions in the analysis of the conditional measurement of two non-orthogonal projectors.
 No satisfactory criterion of the quantum discord, which relies on the analysis of conditional measurement, is formulated in the $d=2$ hidden variables space due to a lack of uniqueness of the dispersion free representation.
 We also make a speculative comment on a "many-worlds interpretation" of hidden variables models to account for the conditional measurement. 
  
\end{abstract}

\maketitle

\section{Introduction}
The phenomenon called quantum discord received much attention in the field of quantum physics recently. It is known that the quantum discord survives even for the separable system without any entanglement\cite{zurek, vedral}. It is then interesting to examine if the hidden
variables models can describe the quantum discord. If the hidden variables models with local realism simulate all the consequences of quantum mechanics except for the (long-ranged) entanglement\cite{epr, bell2, chsh}, it is expected that hidden variables models describe the quantum discord consistently despite the fact that hidden variables models are based on the dispersion free determinism. We discuss this issue on the basis of the explicit model due to Bell in $d=2$~\cite{bell1} which is best known and considered to be a theoretically viable hidden variables model. We also mention that the same conclusion applies to the explicit $d=2$ model by Kochen and Specker~\cite{kochen}.

It may be appropriate to note that the experimental refutation of CHSH inequality does not exclude hidden variables models in $d=2$. The test of CHSH inequality shows that the full contents of quantum mechanics even for a far-apart $d=4$ system cannot be described by separable states only and inseparable states are definitely required~\cite{aspect}. 
The CHSH inequality does not test the quantum description of separable states themselves, which are well described by a direct product of two consistent hidden variables models in $d=2$. The test of hidden variables models in $d=2$ needs to be performed separately, such as the analysis in the present paper. To be specific, we postulate that any physical quantity should have a unique expression in hidden variables space, just as any quantum mechanical quantity has a unique space-time dependence. We then show
the lack of uniqueness of the expression of conditional meaurement in hidden variables space, which we regard as an unsatisfactory feature of the $d=2$ non-contextual hidden varaibles models. The quantum discord, which is based on the conditional measurement, is also influenced by this non-uniqueness when one attempts to describe the criterion of quantum discord in terms of hidden variables models in $d=2$.

\section{Hidden variables models}

For any projector $X$ in noncontextual hidden variables models, we assign a classical number~\cite{bell1, kochen}
\begin{eqnarray}
X \rightarrow X_{\rho}(\omega)
\end{eqnarray}
where $X_{\rho}(\omega)$ assumes its eigenvalues, $X_{\rho}(\omega)=1$ or $X_{\rho}(\omega)=0$, with $\omega\in \Lambda$ standing for hidden variables. The classical number $X_{\rho}(\omega)$
depends on {\em each given pure state} $\rho$ although we often suppress the index $\rho$. 
For a complete set of orthogonal projection operators
$\sum_{k}X_{k}=1$,
we assume the linearity in the sense
\begin{eqnarray}
\sum_{k}X_{k \rho}(\omega)=1.
\end{eqnarray}
We then define
\begin{eqnarray}
x_{\rho}=X_{\rho}^{-1}(1)=\{\omega\in \Lambda : X_{\rho}(\omega)=1\},
\end{eqnarray}
and the probability measure associated with the projection operator $X$ is defined by 
\begin{eqnarray}
\mu[x_{\rho}]\equiv \int_{\Lambda}X_{\rho}(\omega)d\mu(\omega) =\text{Tr}[\rho X].
\end{eqnarray}
The basic assumption in hidden variables models is that one can find $X_{\rho}(\omega)$ and a measure $\mu[x_{\rho}]$ which reproduce the quantum mechanical $\text{Tr}[\rho X]$ for any $X$ and $\rho$. It is known that noncontextual hidden variables models in $d\geq3$ are excluded by Gleason's theorem~\cite{gleason, bell1, beltrametti} and the analysis of Kochen and Specker~\cite{kochen}.

The construction due to Bell in $d=2$ is based on the projector 
\begin{eqnarray}
P_{\vec{m}}=\frac{1}{2}(1+\vec{m}\cdot\vec{\sigma})
\end{eqnarray}
with $|\vec{m}|=1$, and the rule (here, $\frac{1}{2}\geq \omega\geq -\frac{1}{2}$)
\begin{eqnarray}
P_{\vec{m}\psi}(\omega)=\frac{1}{2}[1+\text{sign}(\omega+\frac{1}{2}|\vec{s}\cdot\vec{m}|)\text{sign}(\vec{s}\cdot\vec{m})]
\end{eqnarray}
for the pure state represented by the projector $|\psi\rangle\langle\psi|=\frac{1}{2}(1+\vec{s}\cdot\vec{\sigma})$ with {\em uniform noncontextual} $d\mu(\omega)=d\omega$~\cite{bell2, bell1}. It is shown that the dispersion free $P_{\vec{m}\psi}(\omega)$ itself is not given by any density matrix parameterized by $\vec{s}$ and $\omega$~\cite{beltrametti}. We use the notation of the probability measure $\mu[x_{\psi}]$ for the present case. It is considered that Bell's explicit noncontextual model in $d=2$ is free from the existing no-go theorems in the framework with projectors~\cite{beltrametti, peres}.

\section{Conditional measurement}
 
In quantum mechanics one may first measure a projection operator  $B$~\cite{neumann}. Immediately after the measurement of $B$, one may measure another projector $A$. This operation is allowed even for two non-commuting projectors $[A,B]\neq 0$  
and this operation is called the conditional measurement~\cite{umegaki,davies}.
We now examine Bell's  explicit construction in $d=2$ in connection with the conditional measurement.  

One of the ways to deal with the conditional measurement on the basis of projectors is to define 
\begin{eqnarray}
\rho_{B}\equiv \frac{B\rho B}{\text{Tr}\rho B}, \hspace{1cm} \text{Tr}\rho B \neq 0,
\end{eqnarray}
then the relation 
\begin{eqnarray}
\mu[a_{\rho_{B}}]=\text{Tr}[\rho_{B}A]=\frac{\text{Tr}[\rho BAB]}{\text{Tr}[\rho B]}
\end{eqnarray}
holds as long as the assumed relations (1)-(4) in hidden variables models are valid for any density matrix $\rho$ which includes the density matrix $\rho_{B}$ in (7). This construction of (8) is faithful to the original quantum mechanical definition of the conditional measurement.
In Bell's construction (6), the projected state $\rho_{B}$ corresponds to $|\psi_{B}\rangle\langle\psi_{B}|=B$ in a matrix notation and we have the dispersion free representation (with $A=P_{\vec{m}}$, $B=P_{\vec{n}}$)
\begin{eqnarray}
A_{\psi_{B}}(\omega)=\frac{1}{2}[1+\text{sign}(\omega+\frac{1}{2}|\vec{n}\cdot\vec{m}|)\text{sign}(\vec{n}\cdot\vec{m})]
\end{eqnarray}
which is symmetric in $A$ and $B$, and we obtain the identical expression for $B_{\psi_{A}}(\omega)$.
This expression reproduces the quantum mechanical result (8) if one replaces $\mu[a_{\rho_{B}}]$ by $\mu[a_{\psi_{B}}]$ after averaging over hidden variables.

An alternative way to analyze the conditional measurement is to {\em define} the ratio of averages~\cite{umegaki, davies}
\begin{eqnarray}
\alpha_{B}(A)=\frac{\text{Tr}\rho BAB}{\text{Tr}[\rho B]}, \hspace{1cm} \text{Tr}[\rho B]\neq 0
\end{eqnarray}
as the conditional probability measure of $A$ after the measurement of $B$. 
For the projector in (5), we have
\begin{eqnarray}
P_{\vec{n}}P_{\vec{m}}P_{\vec{n}}=\frac{(1+\vec{n}\cdot\vec{m})}{2}
P_{\vec{n}}
\end{eqnarray}
and $P_{\vec{m}}P_{\vec{n}}P_{\vec{m}}=\frac{1}{2}(1+\vec{n}\cdot\vec{m})P_{\vec{m}}$. It is then natural to define the dispersion free representation
\begin{eqnarray}
(BAB)_{\psi}(\omega)&=&\frac{1}{2}(1+\vec{n}\cdot\vec{m})\nonumber\\
&\times&\frac{1}{2}[1+\text{sign}(\omega+\frac{1}{2}|\vec{s}\cdot\vec{n}|)\text{sign}(\vec{s}\cdot\vec{n})]\nonumber
\end{eqnarray}
or 
\begin{eqnarray}
\frac{(BAB)_{\psi}(\omega)}{\langle B\rangle_{\psi}}&=&\frac{(1+\vec{n}\cdot\vec{m})}{(1+\vec{n}\cdot\vec{s})}\\
&\times&\frac{1}{2}[1+\text{sign}(\omega+\frac{1}{2}|\vec{s}\cdot\vec{n}|)\text{sign}(\vec{s}\cdot\vec{n})]\nonumber
\end{eqnarray}
using (6) for $B_{\psi}(\omega)$ with  $A=P_{\vec{m}}$ and $B=P_{\vec{n}}$. We thus have $\mu[(bab)_{\psi}]=\frac{1}{2}(1+\vec{n}\cdot\vec{m})\mu[b_{\psi}]$.
One then confirms that the conditional measurement is consistently described as
\begin{eqnarray}
\frac{\mu[(bab)_{\psi}]}{\mu[b_{\psi}]}=\frac{\langle BAB\rangle_{\psi}}{\langle B\rangle_{\psi}}
=\frac{\langle ABA\rangle_{\psi}}{\langle A\rangle_{\psi}}=\frac{(1+\vec{n}\cdot\vec{m})}{2}.
\end{eqnarray}
It is important to
note that $(BAB)_{\psi}(\omega)\neq B_{\psi}(\omega)A_{\psi}(\omega)B_{\psi}(\omega)$, for example, where the left-hand side stands for a single positive operator (which uses the quantum mechanical product rule) while the right-hand side stands for a product of projectors (which implies $B_{\psi}(\omega)A_{\psi}(\omega)B_{\psi}(\omega)=A_{\psi}(\omega)B_{\psi}(\omega)=A_{\psi}(\omega)B_{\psi}(\omega)A_{\psi}(\omega)$).

This specific example in (13) shows that the conditional measurement in hidden variables models~\cite{bell1} does not follow the classical conditional probability rule
\begin{eqnarray}
\frac{\mu[(bab)_{\psi}]}{\mu[b_{\psi}]} \neq \frac{\mu[a_{\rho}\cap b_{\rho}]}{\mu[ b_{\rho}]}
\end{eqnarray}
for general non-commuting $A$ and $B$, despite the fact that hidden variables models are based on the dispersion free determinism. If one assumes the classical conditional probability rule on the right-hand side of (14) for general state $\rho$, the relation (13) cannot hold for non-commuting $A$ and $B$~\cite{malley2004}. 

We now recognize that the expression (9) and the expression (12)  lead to two conflicting dispersion free representations in  hidden variables space parameterized by $\omega$ for the same quantum mechanical object $\text{Tr}[\rho BAB]/\text{Tr}[\rho B]$, although both of them reproduce the same quantum mechanical result after averaging over hidden variables. 
This conflict between (9) and (12) is analogous to the case of a sum of two non-orthogonal projectors.
One may consider a linear combination of two non-collinear projectors in (5)
\begin{eqnarray}
E=\lambda P_{\vec{n}}+ (1-\lambda) P_{\vec{m}}, \ \ \ \  0< \lambda< 1,
\end{eqnarray}
which satisfies $0<E<1$. If one assumes that the dispersion free representation due to Bell is applied to all the operators in (15) separately, one obtains
\begin{eqnarray}
E_{\psi}(\omega)=\lambda P_{\vec{n}\psi}(\omega)+ (1-\lambda) P_{\vec{m}\psi}(\omega), \ \ \ \ 0< \lambda< 1,
\end{eqnarray}
but this is not satisfied by the positive operator  $1>E_{\psi}(\omega)>0$ on the left-hand side in the domain of the hidden variables space
with $P_{\vec{n}\psi}(\omega)=P_{\vec{m}\psi}(\omega)=0$ (or with
$P_{\vec{n}\psi}(\omega)=P_{\vec{m}\psi}(\omega)=1$). This shows that Bell's construction has an ambiguity in representing the same operator, namely, the left and right-hand sides of (15), although it reproduces the result of quantum mechanics $\langle E\rangle_{\psi}=\lambda\langle  P_{\vec{n}}\rangle_{\psi}+ (1-\lambda) \langle P_{\vec{m}}\rangle_{\psi}$ implied by (15) after averaging over hidden variables.

The conflict in (16) is essentially the original no-go theorem of von Neumann~\cite{neumann} against noncontextual hidden variables models, and its resolution is well known. One does not assign a physical significance to two incompatible operators simultaneously in hidden variables space~\cite{bell1}. One now encounters another conflict between (9) and (12) arising from non-orthogonal projectors in the analysis of the conditional measurement. 

We examine the conflict between (9) and (12) in more detail.
From a point of view of the dual structure of operator and state $(O,\rho)$ in quantum mechanics, these two approaches are related; an extra quantum mechanical operation is included in each case, $(A,B\rho B)$ or $(BAB,\rho)$ respectively, before moving to hidden variables models. These two are obviously equivalent in quantum mechanics (or in any trace representation with density matrix), but they are quite different in Bell's construction due to  the lack of definite associative properties of various operations. 

The hidden variables prescription, which does not translate the combination $\text{Tr}[\rho A]$ directly, contains ambiguity
as is seen in $\psi^{\dagger}A\psi=(A\psi)^{\dagger}1(A\psi)$; in the context of hidden variables models, the first expression implies a dispersion free representation of $A$ for $\psi$, while the second 
implies a dispersion free representation of unit operator for $(A\psi)$. We thus have to specify both of the {\em operator} and {\em state } which we want to  express by a dispersion free representation. One may choose the representation by a physical picture. In our examples of (9) and (12), the expression (9) has a more direct physical meaning as a dispersion free representation of $A$ immediately after the measurement of $B$. Bell's construction works for general positive operators such as $BAB$ in (12) in addition to projectors, but the expression (12) gives a less direct physical picture of a dispersion free representation of $A$ immediately after the measurement of $B$. Incidentally, this ambiguity  disappears in the case of two orthogonal projectors $AB=0$.  
 
Each prescription in (9) and (12) has its own unsatisfactory features. The construction $\mu[a_{\rho_{B}}]$ (8) and also the expression (9) rely directly on the notion of reduction of states caused by a measurement of $B$, which is a characteristically quantum mechanical notion foreign to deterministic realism. One of the motivations of hidden variables models is to avoid the sudden reduction of states. In fact, the replacement of the dispersion full $\langle\psi|P_{\vec{m}}|\psi\rangle$ by the dispersion free $P_{\vec{m}\psi}(\omega)$ in (6) is achieved by avoiding the reduction. It may thus be desirable to avoid the direct use of reduction in hidden variables models. In comparison, the expression (10) with (12) uses the same state before and after the measurement of $B$ and thus avoids the issue of reduction (by using the quantum mechanical product $BAB$ instead), but now
one has to explain the state preparation, namely, how to produce a desired state such as in (6) by an experimental procedure without referring to reduction (or its possible counter part in hidden variables models, which is not specified in Bell's construction).

  One may conclude that no satisfactory dispersion free description of the conditional measurement exists in Bell's construction besides the lack of uniqueness of the expression in hidden variables space.
(An interesting example is given by the measurement of $A$ immediately after the measurement of $A$. The prescription in (9) gives an $\omega$ independent unit representation, while the formula (10) with (12) gives $A_{\psi}(\omega)/\int A_{\psi}(\omega)d\omega$ which has the same $\omega$ dependence as the first measurement.)  

We mention that essentially the same conclusion holds in connection with the conditional measurement in the explicit $d=2$ hidden variables model in~\cite{kochen} also.

\section{Quantum Discord}

The definition of the quantum discord for a two-partite system described by $\rho_{XY}$ is given by the difference of the quantum conditional entropy
\begin{eqnarray}
\sum_{j}p_{j}S(\rho_{Y|\Pi^{X}_{j}})
\end{eqnarray}
and the quantity which formally corresponds to the conditional entropy $S(X,Y)-S(X)$, namely,~\cite{zurek}\cite{vedral}
\begin{eqnarray}
D=\sum_{j}p_{j}S(\rho_{Y|\Pi^{X}_{j}})
-[S(X,Y)-S(X)].
\end{eqnarray}
Here we choose a suitable set of orthogonal projectors $\Pi^{X}_{i}\Pi^{X}_{j}=\Pi^{X}_{j}\Pi^{X}_{i}=\delta_{i,j}\Pi^{X}_{j},\ \ \sum_{j}\Pi^{X}_{j}=1$,
and defined 
\begin{eqnarray}
\rho_{Y|\Pi^{X}_{j}}=\frac{\text{Tr}_{X}[(\Pi^{X}_{j}\otimes 1)\rho_{XY}(\Pi^{X}_{j}\otimes 1)]}{p_{j}}
\end{eqnarray}
with $p_{j}=\text{Tr}[(\Pi^{X}_{j}\otimes 1)\rho_{XY}]$. The quantum discord $D$ is actually defined at the minimum of the first term in (18) with respect to all the possible choices of the set $\{\Pi^{X}_{j}\}$. 

We have the relation 
\begin{eqnarray}
&&\sum_{j}p_{j}S(\rho_{Y|\Pi^{X}_{j}})\nonumber\\
&=&\sum_{j}p_{j}\log p_{j}-\sum_{j}\text{Tr}_{Y}\{\text{Tr}_{X}[(\Pi^{X}_{j}\otimes 1)\rho_{XY}(\Pi^{X}_{j}\otimes 1)]\nonumber\\
&&\times \log \text{Tr}_{X}[(\Pi^{X}_{j}\otimes 1)\rho_{XY}(\Pi^{X}_{j}\otimes 1)]\}.
\end{eqnarray}
The following general properties of the quantum discord are known:~\cite{zurek, wehrl, datta}
\begin{eqnarray}
&& D=0 \Longleftrightarrow \rho_{XY}=\sum_{j}\Pi^{X}_{j}\rho_{XY}\Pi^{X}_{j}=\sum_{j}p_{j}\Pi^{X}_{j}\otimes\rho^{Y}_{j},\nonumber\\
&&0\leq D\leq S(\rho_{X}).
\end{eqnarray}
See also Refs.\cite{vedral2, datta2} for more recent related analyses.

We now analyze the {\em vanishing condition} of the quantum discord, namely, the first condition in (21).
We note a similarity between the reduction in (7) and the first condition in (21), and it is necessary to show 
\begin{eqnarray}
\text{Tr}_{X}A_{X}\rho_{XY}=\sum_{j}\text{Tr}_{X}A_{X}\Pi^{X}_{j}\rho_{XY}\Pi^{X}_{j}
\end{eqnarray}
for any projector $A_{X}$ to establish the first condition in (21). We thus have to deal with general positive operators 
, $\Pi^{X}_{j}A_{X}\Pi^{X}_{j}$, to discuss the criterion of the vanishing quantum discord. In the case of separable state $\rho_{XY}=\sum_{k}w_{k}\rho^{(k)}_{X}\otimes\rho^{(k)}_{Y}$, which is relevant in the present context if one accepts the absence of entanglement in noncontextual hidden variables models with local realism,
the condition (22) becomes
\begin{eqnarray}
&&\sum_{k}w_{k}[\text{Tr}_{X}A_{X}\rho^{(k)}_{X}]\rho^{(k)}_{Y}
\nonumber\\
&&=\sum_{k}w_{k}\sum_{j}[\text{Tr}_{X}A_{X}\Pi^{X}_{j}\rho^{(k)}_{X}\Pi^{X}_{j}]\rho^{(k)}_{Y}.
\end{eqnarray}
In the explicit construction due to Bell~\cite{bell1} one can simulate the right-hand side of (23) (by using a quantum mechanical product rule before moving to hidden variables models), but two different prescriptions lead to conflicting expressions in hidden variables space as is shown in (9) and (12). Quantum mechanically equivalent expressions lead to quite different dispersion free expressions in Bell's construction, and there appears to be no clear physical criterion to resolve the ambiguity.
One may thus conclude that the description  of the quantum discord in hidden variables space is ill-defined in  Bell's construction (and probably in any noncontextual hidden variables models without density matrix representation). 

One may wonder why the conditional measurement appears in (23) while the simple average $\text{Tr}[\Pi^{X}_{j}\rho_{X}]$, which is simulated by Bell's construction without referring to general positive operators, appears 
in the expression such as (18). The reason is that one evaluates a candidate of the quantum discord for a given state $\rho_{X}$  and $\{\Pi^{X}_{j}\}$ in (18) (the relation $\text{Tr}[\Pi^{X}_{j}\rho_{X}]=\text{Tr}[\Pi^{X}_{j}\sum_{k}\Pi^{X}_{k}\rho_{X}\Pi^{X}_{k}]$ does not provide any constraint on $\rho_{X}$), while (23) is a {\em condition} on the state $\rho_{X}$ and $\{\Pi^{X}_{j}\}$ and thus one needs to examine the full properties of states after projection.
 
In passing, we comment on the scheme with positive operator valued measures (POVMs), although in the present paper we work with projection operators for which $d=2$ noncontextual hidden variables models are consistently defined. For the scheme with POVMs where allowed positive operators in the resolution of unity $\sum_{k}E_{k}=1$ are increased from the projectors in (2) and the linearity of the probability measure is extended to $v(\sum_{k}E_{k})=\sum_{k}v(E_{k})=1$,  noncontextual hidden variables models are excluded even in $d=2$~\cite{busch}. 
  
\section{Discussion} 

We briefly discuss the implications of the classical conditional probability rule in the conditional measurement
\begin{eqnarray}
\frac{\mu[a_{\rho}\cap b_{\rho}]}{\mu[ b_{\rho}]}
=\frac{\text{Tr}[\rho BAB]}{\text{Tr}[\rho B]}
\end{eqnarray}
in general noncontextual hidden variables models.
If (24) is assumed to hold for general $A$ and $B$, and thus the relation with $A$ and $B$ interchanged also, one has 
\begin{eqnarray}
\text{Tr}[\rho BAB]=\text{Tr}[\rho ABA].
\end{eqnarray}

If the relation (25) is assumed to be valid for any $\rho$, it is known that (25) leads to the constraint on operators $[A,B]=0$~\cite{malley}. 
  
On the other hand, if (24) is understood as a {\em constraint on the state} $\rho$ for arbitrary $A$ and $B$, which is natural in the context of quantum discord, it leads to $\rho\propto 1$. To see this, we consider 
$\text{Tr}_{X}[{\cal P}^{X}_{l}\Pi^{X}_{j}\rho_{X}\Pi^{X}_{j}]$
with a suitable complete set of orthogonal projectors of the $X$ system
$\sum_{l}{\cal P}^{X}_{l}=1$.
The property (25) implies
\begin{eqnarray}
\text{Tr}_{X}[{\cal P}^{X}_{l}\Pi^{X}_{j}\rho_{X}\Pi^{X}_{j}]
=\text{Tr}_{X}[\Pi^{X}_{j}{\cal P}^{X}_{l}\rho_{X}{\cal P}^{X}_{l}],
\end{eqnarray}
which, after summing over $l$, implies a weaker relation
$\text{Tr}_{X}\Pi^{X}_{j}\rho_{X}
=\text{Tr}_{X}\Pi^{X}_{j}\sum_{l}[{\cal P}^{X}_{l}]\rho_{X}[{\cal P}^{X}_{l}].$
This last relation valid for any $\Pi^{X}_{j}$ leads to
$\rho_{X}=
\sum_{l}{\cal P}^{X}_{l}\rho_{X}{\cal P}^{X}_{l}$,
which holds for any set $\{P^{X}_{l}\}$ in turn implies a specific (completely degenerate) mixed state $\rho_{X}\propto 1$. For the separable state $\rho_{XY}=\sum_{k}w_{k}\rho^{(k)}_{X}\otimes\rho^{(k)}_{Y}$ which is relevant in the present context of hidden variables models, the classical conditional probability rule in the party $X$  implies $\rho_{XY}=\sum_{k}w_{k}{1}_{X}\otimes\rho^{(k)}_{Y}$ and thus gives a vanishing quantum discord. The classical conditional probability rule (24), if imposed on noncontextual hidden variables models, thus eliminates all the quantum mechanical properties. 
 The left-hand side of (24) negates the crucial notion of reduction in quantum mechanics, as is seen by the fact that $a_{\rho}$ and $b_{\rho}$ in $\mu[a_{\rho}\cap b_{\rho}]$ are defined by the same original state $\rho$ although $\mu[a_{\rho}\cap b_{\rho}]$ is divided by $\mu[ b_{\rho}]$. 
\\

The quantum discord which is based on the conditional measurement
provides a further test of the basic aspects of quantum mechanics such as possible hidden variables models.  On the basis of Bell's explicit construction in $d=2$ which has no density matrix representation, it was pointed out
that the description of the criterion of quantum discord in hidden variables space is ill-defined. The same conclusion applies to the $d=2$ model by Kochen and Specker. The hidden variables models are useful to deepen
our understanding of the "non-classical" properties of quantum mechanics, and although in the framework of non-contextual hidden variables models in $d=2$, the "quantumness" of quantum discord is traced to the reduction of states, as is seen by the comparison of (7) with (21), in contrast to the locality which is emphasized in the analysis of entanglement.

We finally add a speculative comment on a way to avoid the discrepancy between the expressions (9) and (12) in hidden variables space. One may unify these two expressions as
\begin{eqnarray}
A_{\psi_{B}}(\omega^{\prime})B_{\psi}(\omega)/\int d\omega B_{\psi}(\omega)
\end{eqnarray}
and later integrate over $\omega$ and $\omega^{\prime}$ independently; if one integrates over $\omega$ first one obtains (9), while  if one integrates over $\omega^{\prime}$ first one obtains (12).  
Namely, one may assume that each measurement opens up a new hidden variables space, which effectively incorporates the notion of reduction, and this procedure is somewhat analogous to the "many-worlds interpretation"~\cite{everett}.
An analysis of entire hidden variables models in the scheme analogous to the many-worlds interpretation is an interesting subject.
\\

I thank C.H. Oh, L.C. Kwek and S.X. Yu for informative discussions at National University of Singapore.


\begin{thebibliography}{99}
\bibitem{zurek}
H. Ollivier and W. H.  Zurek, Phys. Rev. Lett. {\bf 88}, 017901 (2002) .
\bibitem{vedral}
L. Henderson and V. Vedral, J. Phys. A{\bf 34}, 6899 (2001).
\bibitem{epr}
A. Einstein, B. Podolsky and N. Rosen, Phys. Rev. {\bf 47}, 777 (1935).
\bibitem{bell2}
J. S. Bell, Physics {\bf 1}, 195 (1965).
\bibitem{chsh}
J. F. Clauser, M. A. Horne, A. Shimony and R. A. Holt, Phys. Rev. Lett. 
{\bf 23},  888 (1969).
\bibitem{bell1}
J. S. Bell, Rev. Mod. Phys. {\bf 38}, 447 (1966).
\bibitem{kochen}
S. Kochen and E. P. Specker, J. Math. Mech. {\bf 17}, 59 (1967).
\bibitem{aspect}
A. Aspect, J. Dalibard and G. Roger, Phys. Rev. Lett. {\bf 49},
1804 (1982).
\bibitem{gleason}
A. M. Gleason, J. Math. Mech. {\bf 6}, 885 (1957).
\bibitem{beltrametti}
E. G. Beltrametti and G. Gassinelli, {\em The Logic of Quantum 
Mechanics}, (Addison-Wesley Pub., 1981).
\bibitem{peres}
A. Peres, {\em Quantum Theory: Concepts and Methods }, (Kluwer Academic Pub., 1995).
\bibitem{neumann}
J. von Neumann, {\em Mathematical Foundations of Quantum Mechanics}
(Princeton Univ. Press, 1955).
\bibitem{umegaki}
H. Umegaki, Tohoku Math. J. {\bf 6}, 177 (1954).
\bibitem{davies}
E. B. Davies and J. T. Lewis, Comm. Math. Phys. {\bf 17}, 239 (1970).  
\bibitem{malley2004}
The complications associated with the analysis of the conditional measurement in hidden variables models have been discussed from a quite different point of view in\\
J. D. Malley, Phys. Rev. A{\bf 69}, 022118 (2004),\\
K. Nagata, Phys. Rev. A{\bf 73}, 066101 (2006),\\
J. D. Malley and A. Fine, Phys. Rev. A{\bf 73}, 066102 (2006).\\
Our analysis which is based on concrete models in~\cite{bell1, kochen} is more explicit and elementary. We believe that our analysis pinpoints the origin of the difficulty associated with the analysis of the conditional measurement in hidden variables models.
\bibitem{wehrl}
A. Wehrl, Rev. Mod. Phys. {\bf 50}, 221 (1978).
\bibitem{datta}
A. Datta, {\em Ph. D. Thesis}, University of New Mexico (2008), arXiv:0807.4490v1.
\bibitem{vedral2}
B. Dakic, V. Vedral and C. Brukner, Phys. Rev. Lett. {\bf 105}, 190502 (2010).
\bibitem{datta2}
A. Datta, arXiv:1003.5256v1[quant-ph].
\bibitem{busch}
P. Busch,  Phys. Rev. Lett. {\bf 91}, 120403 (2003).
\bibitem{malley}
J. D. Malley and A. Fine, Phys. Lett. A {\bf 347}, 51 (2005). 
\bibitem{everett}
B.S. DeWitt and R.N. Graham, eds., {\em The Many-Worlds Interpretation of Quantum Mechanics}, Princeton Series in Physics, (Princeton University Press, 1973).
\end{thebibliography}
\end{document}